\documentclass[11pt]{article}
\usepackage{psfig}

\newlength{\figwidth}
\title{
\vspace{-1cm}
\begin{flushright}
\normalsize  RCNP-Th01010
\end{flushright}
\vspace{1cm}
SU(3) lattice QCD study for octet and decuplet baryon spectra%
\thanks{Talk presented by N.N. at Symposium on Hadrons and Nuclei,
            20-22 Feb. 2001, Seoul, Korea.} }

\author{
N.~Nakajima$^a$, H.~Matsufuru$^a$, Y.~Nemoto$^b$ and H.~Suganuma$^c$\\[1.2em]
   {\it $^a$Research Center for Nuclear Physics, Osaka University,
            Ibaraki 567-0047, Japan} \\
   {\it $^b$Yukawa Institute for Theoretical Physics, Kyoto University,
            Kyoto 606-8502, Japan}\\
   {\it $^d$Faculty of Science, Tokyo Institute of Technology,
            Tokyo 152-8551, Japan} }

\date{May 8, 2001}

\begin{document}

\maketitle

\begin{abstract}
The spectra of octet and decuplet baryons are studied 
using $SU(3)$ lattice QCD at the quenched level.
As an implementation to reduce the statistical fluctuation, 
we employ the anisotropic lattice with $O(a)$ improved quark action.
In relation to $\Lambda(1405)$, 
we measure also the mass of the $SU(3)$ flavor-singlet negative-parity 
baryon, which is described as a three quark state in the quenched lattice QCD, 
and its lowest mass is measured about 1.6 GeV. 
Since the experimentally observed negative-parity baryon 
$\Lambda(1405)$ is much lighter than 1.6 GeV, 
$\Lambda(1405)$ may include a large component of a $N \bar K$ 
bound state rather than the three quark state. 
The mass splitting between the octet and the decuplet baryons are
also discussed in terms of the current quark mass.
\end{abstract}

\section{Introduction}
  \label{sec:Introduction}

The lattice QCD simulation has become a powerful method to
investigate hadron properties directly based on QCD.
The hadron spectroscopy in the quenched approximation, i.e.
without the dynamical quark effect, has been almost established,
and reproduces experimental values of the low-lying hadron
masses within 10 \% deviation \cite{CPPACS}.
Detailed investigation with dynamical quarks will give us
an insight on the dynamical quark effect on hadron properties
\cite{FullQCD}.
With these situations, we now have a stage to proceed into more
extensive and systematic studies of hadron structure
in lattice QCD, and compare the lattice results with the model 
analysis or the phenomenological approaches such as the potential models
\cite{Suganuma}.
The latter point of view would give us clearer physical picture of
extracted information from lattice simulations,
and may be useful to extract the physical quantities beyond 
the lattice QCD applicability.
Furthermore, comparing hadron properties in the lattice QCD with   
the potential model analysis using the potential derived from lattice QCD, 
one can verify the applicability of these approaches in
a self-consistent manner.
This is our motivation of lattice calculation of the static
three quark potential, which is responsible to the baryon
properties \cite{Takahashi}.

Compared with the ground state hadrons, excited state hadrons are
far from established.
The purpose of our present study is to perform detailed investigation
of excited states as well as the ground state hadrons.
Among them, in this paper, we focus on two important subjects:
\begin{description}

\item[\rm (a)] Origin of Octet-Decuplet baryon mass splitting. 

There have been proposed several models to explain the low-lying
hadron spectrum.
As one of celebrated models, the nonrelativistic quark model
explains the $N$-$\Delta$ mass splitting with one-gluon-exchange
(OGE) interaction and results in \cite{IK79,OY81}
\begin{equation}
M_{\Delta}-M_{N} \propto \sum_{i<j}\frac{1}{M_i M_j} ,
\label{eq:OGEmass}
\end{equation}
where $M_i$ is the {\it constituent} quark mass.
This implies mass difference between octet and decuplet baryons
decreases with increasing quark mass.
In lattice QCD simulation, it is possible to change the {\it current}
quark mass through the hopping parameter $\kappa$.
Then we compare the obtained mass splitting with
the form (1), although the relation between current and constituent 
quark masses is not so clear.

\item[\rm (b)] Structure of $\Lambda(1405)$ and other negative parity 
baryons.

The detailed lattice study of the negative-parity baryons has been 
started rather recently \cite{NPbaryons}.
Among the negative-parity baryons, we pay much attention to
$\Lambda(1405)$ with $J^{P}=1/2^-$, since its structure in terms of
the constituent quark picture is not well understood.
There are interesting two possibilities proposed for $\Lambda(1405)$,  
an $SU(3)_f$ singlet state $(qqq)$ and a $N\bar{K}$ bound state as 
($qqq$-$q\bar{q}$).
In this paper, we investigate the flavor-singlet baryon spectrum with 
spin $1/2$ and negative parity in lattice QCD, and compare it with 
$\Lambda(1405)$.  
The second possibility as the $N \bar K$ bound state 
is now in progress at the quenched level, 
where quark-antiquark pair creation (strictly speaking, dynamical
quark loop effect) is absent and then the quark-level 
constitution in hadrons is definitely clear in the simulation. 

\end{description}

In lattice QCD simulations, not only the quark mass,
we can also change the number of dynamical quark flavor.
This enables us to extract the quark loop effect on the hadron 
properties.
However, the flavor-number dependence does not seem significant for 
the low-lying hadron spectrum.  Then, we start with calculations 
at the quenched level (with no dynamical flavor). 
Before proceeding to the numerical calculation, however,
we should describe a technical problem and equipment to
circumvent it.

The difficulty of extracting the excited state masses
lies in the rapid growth of the statistical fluctuations
in the correlation functions. 
In the practical simulation for the correlation function, 
it is hard to identify the reliable range of its Euclidean temporal distance 
where the relevant information is kept without suffering from the 
large statistical noise.
To overcome this problem, we adopt the anisotropic lattice,
on which the temporal lattice spacing $a_{\tau}$ is finer than
the spatial one, $a_{\sigma}$ \cite{Kar82}.
Detailed information in the temporal direction makes
these analyses as the mass measurements extensively easy.
This approach is especially efficient for the heavy particle correlators, 
such as excited states and glueballs, 
for which the noises grow rapidly against the signals. 
On the other hand, introduction of anisotropy requires us
additional effort in the numerical simulations.
Due to the quantum effect, the renormalized anisotropy 
$\xi\equiv a_{\sigma}/a_{\tau}$ differs from the bare one 
in general, and at the first stage of the
simulation one need to tune the bare anisotropy so that
the quark field retains the same renormalized anisotropy
as the gluon field.

This paper is organized as follows.
In the next section, we briefly summarize the anisotropic lattice,
especially the quark action.
Then, we describe the numerical simulation
and discuss on the obtained result.
The last section gives our summary and outlook.

\section{Anisotropic lattice}
 \label{sec:action}

The anisotropic lattice has become an extensively useful tool
in the lattice QCD simulations.
In addition to aforementioned advantage, fine temporal
resolution is particularly significant to extract the
information of hadron correlators at finite temperature.
Another advantage is that it can treat the relatively heavy quark
without introducing the effective theoretical approaches.
This feature is particularly suited for the study of
charmonium and charmed hadrons.
Although this is not a subject of this paper, this is one of
reasons we adopt the anisotropic lattice.
In this work, we treat the hadron correlators, and hence we focus here 
on the quark action on the anisotropic lattice.

As the quark action on the anisotropic lattice, we adopt the
$O(a)$ improved Wilson action \cite{Kla99}.
The construction of the action is along the program of Fermilab
formulation \cite{EKM97}.
Their treatment incorporates the full quark-mass dependence in the
construction of lattice quark action, so that the
quark mass region with $m_q\simeq a^{-1}$ is also available
without large $O(ma)$ uncertainty.
The Fermilab approach is naturally generalized to the anisotropic
lattice.
There is some arbitrariness in choosing the parameterization of action,
and we use the form proposed in \cite{Ume01,Aniso01a}.
In \cite{Aniso01a}, the advantages of this form are discussed in
detail.
The quark action is written in the following form:
\begin{equation}
S_F = \sum_{x,y} \bar{\psi}(x) K(x,y) \psi(y),
\vspace{-0.5cm}
\end{equation}
\begin{eqnarray}
 K(x,y) \, = \, \delta_{x,y}
  &-& \kappa_{\tau} \left\{
     (1-\gamma_4) U_4(x) \delta_{x+\hat{4},y}
     (1+\gamma_4) U_4^{\dag}(x-\hat{4}) \delta_{x-\hat{4},y} \right\}
 \nonumber \\
  &-& \kappa_{\sigma} \sum_i \left\{
     (r-\gamma_i) U_i(x) \delta_{x+\hat{i},y}
     (r+\gamma_i) U_i^{\dag}(x-\hat{i}) \delta_{x-\hat{i},y} \right\}
 \label{eq:action}  \\
  &-& \kappa_{\sigma} c_E \sum_i \sigma_{i4} F_{i4} \delta_{x,y}
  - r \kappa_{\sigma} c_B \sum_{ij} \frac{1}{2}
                   \sigma_{ij} F_{ij} \delta_{x,y}.
 \nonumber
\end{eqnarray}
The gluon field is represented with the link variable
$U_{\mu}\simeq \exp(-iga_{\mu}A_{\mu})$,
and $\psi$ denotes the anticommuting quark field.
The spatial and the temporal hopping parameters, $\kappa_{\sigma}$
and $\kappa_{\tau}$, respectively, are related to the bare quark mass
$m_0$ and the bare anisotropy parameter $\gamma_F$ as 
\begin{equation}
\kappa_{\sigma} = 1/2(m_0 + \gamma_F + 3r), \hspace{1cm} 
\kappa_{\tau} = \gamma_F \kappa_{\sigma},
\end{equation}
where $m_0$ is measured in the spatial lattice unit.
The value of the Wilson parameter $r$ is set as $r=1/\xi$.
The coefficients $c_E$ and $c_B$ in the clover terms are introduced
to eliminate the $O(a)$ error induced by the Wilson term,
and coincide with unity at the tree level.

We apply the mean-field improvement
which reduces large contributions from the tadpole diagrams
to the renormalization.
This is achieved by replacing the link variable as
$U \rightarrow U/u_0$, with the mean-field value of the link variable,
$u_0$.
On the anisotropic lattice, two mean-field values
$u_{\sigma}$ and $u_{\tau}$ are defined for the spatial and
the temporal link variables, respectively.
We employ the definition of the mean-field value
through the average of $U$ in the Landau gauge.
With $u_{\sigma}$ and $u_{\tau}$, the mean-field improved values 
of the clover coefficients at the tree level are expressed as 
$c_E  = 1/ u_{\tau}^2 u_{\sigma}$ and $c_B  = 1 / u_{\sigma}^3$. 
The anisotropy parameter is related to the improved anisotropy
$\tilde{\gamma}_F$ as
$\gamma_F = \tilde{\gamma}_F \cdot u_{\sigma}/u_{\tau}$.
It is convenient to define $\kappa$, as
\begin{equation}
\frac{1}{\kappa} = \frac{1}{\kappa_{\sigma}u_{\sigma}}
  - 2 (\tilde{\gamma}_F + 3r -4)
 \hspace{1cm}
 (=2(m_0+4)).
\end{equation}
For the light quark systems, the extrapolation to the chiral limit
is performed in $1/\kappa$.

In practical simulations, the anisotropy parameter $\gamma_F$ 
should be tuned so that the fermionic anisotropy $\xi_F$ defined
with fermionic observable coincides with the gauge field anisotropy.
This is called as ``calibration''.
Several procedures have been used for the calibration.
In this paper, we set the value of $\gamma_F$ using
the dispersion relation of the pseudoscalar and the vector mesons.
We assume that the meson field is described with the 
lattice Klein-Gordon equation, 
\begin{equation}
 S = \frac{1}{2\xi_F} \sum_{x} \phi^{\dag}(x) 
   \left[ - \xi_F^2 D_4^2 - \vec{D}^2 + m_0^2 \right] \phi(x),
\end{equation}
with $D_{\mu}$ the lattice covariant derivative.
The free meson field satisfies the dispersion relation, 
\begin{equation}
 \cosh E(\vec{p}) - \cosh E(0) = \vec{p}^2 / 2 \xi_F^2.
\end{equation}
This relation is used to define the fermionic anisotropy $\xi_F$
\cite{Ume01}.

\section{Numerical Results}
 \label{sec:result}

\paragraph{Lattice Setup}

The SU(3) lattice QCD simulations are performed on an anisotropic lattice
of the size $12^3\times 96$ with anisotropy $\xi=4$,
at the quenched level.
As the gauge field action, we adopt the anisotropic Wilson action with 
the parameters $(\beta, \gamma_G)=(5.75, 3.072)$,
determined by Klassen so as to give the renormalized anisotropy
$\xi=4$ within 1 \% uncertainty \cite{Kla98}.
At these parameters, the lattice cutoff
defined by setting the string tension $\sqrt{\sigma}$ 
to be 427 MeV is found to be $a_{\sigma}^{-1}(\sqrt{\sigma})\simeq 1.0$ GeV.
The gauge configuration is fixed to the Coulomb gauge,
which is convenient to smear the quark propagators by extending 
the quark source spatially on a time slice.  

The mean-field values are determined on the lattice of half size
in the temporal direction, $12^3\times 48$, at the same 
$(\beta, \gamma_G)$.
We fix these gauge configurations to the Landau gauge and 
determine the mean-field values $u_{\sigma}$ and $u_{\tau}$
self-consistently as described in \cite{Ume01}.
They result in $u_{\sigma} = 0.7620(2)$ and 
$u_{\sigma} = 0.9871$ (error is less than the last digit).

\paragraph{Calibration}
The quark field calibration is performed along the course described
in the last section.
The pseudoscalar and the vector meson correlators are calculated
with momentum $\vec{p}=0$, $2\pi/16$ and $2\cdot 2\pi/16$.
We use the meson operators listed in Table~\ref{tab:operator}
and a standard procedure to extract the meson energy.

\begin{figure}[tb]
\centerline{
\leavevmode\psfig{file=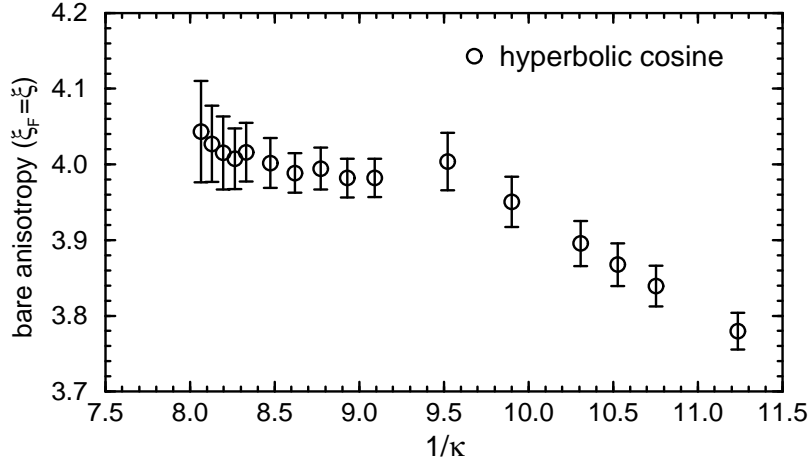,width=0.9\figwidth}}
\caption{Results of the calibration in an anisotropic lattice. 
The value of $\tilde{\gamma}_F$
at which $\xi_F=\xi$ holds is determined for each value of $\kappa$.}
\label{fig:calibration}
\end{figure}

Figure~\ref{fig:calibration} shows the result of the calibration.
The left three values of $\kappa$ correspond to the 
quark masses around the strange quark mass.
We use these three values of $\kappa$ for the following analysis
of hadron spectroscopy.
As is clearly observed in Fig.~\ref{fig:calibration},
in the light quark region ($m_q \simeq m_s$), one can set the
bare anisotropy $\tilde{\gamma}_F=4$.
In Table~\ref{tab:mesonmass}, we list the values of $\kappa$, 
the pseudoscalar and the vector meson masses 
for the degenerate quark case.
The chiral extrapolation is carried out linearly in $1/\kappa$,
and results in the critical hopping parameter as $\kappa_c=0.12637(2)$.

\begin{table}[tb]
\vspace*{0.8cm}
\begin{center}
\begin{tabular}{ccc}
\hline
 \hspace{1cm} $\kappa$  \hspace{1cm} &
 \hspace{1cm}  $m_{PS}$ \hspace{1cm} &
 \hspace{1cm} $m_{V}$  \hspace{1cm} \\
\hline
 0.124 & 0.1504( 6) & 0.2240(22) \\
 0.123 & 0.2036( 6) & 0.2602(13) \\
 0.122 & 0.2499( 5) & 0.2966( 9) \\
\hline
\end{tabular}
\end{center}
\caption{
The hopping parameters used in the spectroscopy
and the PS and V meson masses for the degenerate quark case.}
\label{tab:mesonmass}
\end{table}

\begin{table}[tb]
\small
\begin{tabular}{ccc}
\hline
Meson &
  Pseudoscalar \hspace{-1.0cm} & $M(K)   = \bar{s} \gamma_5 u$ \\
& Vector       & $M_k(K^*) = \bar{s} \gamma_k u$ \\
\hline
Baryon &
  Octet &
  $B_{\alpha}(\Sigma^0) = (C\gamma_5)_{\beta\gamma}[
      u_{\alpha}(d_{\beta}s_{\gamma} - s_{\beta}d_{\gamma}) 
    - d_{\alpha}(s_{\beta}u_{\gamma} - u_{\beta}s_{\gamma}) ]$ \\
&  Octet ($\Lambda$) &
  $B_{\alpha}(\Lambda) = (C\gamma_5)_{\beta\gamma}[
      u_{\alpha}(d_{\beta}s_{\gamma} - s_{\beta}d_{\gamma}) 
    + d_{\alpha}(s_{\beta}u_{\gamma} - u_{\beta}s_{\gamma})
   -2 s_{\alpha}(u_{\beta}d_{\gamma} - d_{\beta}u_{\gamma}) ]$ \\
&  Singlet  &
  $B_{\alpha}(\Lambda_1) = (C\gamma_5)_{\beta\gamma}[
      u_{\alpha}(d_{\beta}s_{\gamma} - s_{\beta}d_{\gamma}) 
    + d_{\alpha}(s_{\beta}u_{\gamma} - u_{\beta}s_{\gamma})
    + s_{\alpha}(u_{\beta}d_{\gamma} - d_{\beta}u_{\gamma}) ]$ \\
&  Decuplet &
  $B_{\alpha k}(\Sigma^{*0}) = (C\gamma_k)_{\beta\gamma}[
      u_{\alpha}(d_{\beta}s_{\gamma} + s_{\beta}d_{\gamma}) 
    + d_{\alpha}(s_{\beta}u_{\gamma} + u_{\beta}s_{\gamma})
    + s_{\alpha}(u_{\beta}d_{\gamma} + d_{\beta}u_{\gamma}) ]$ \\
\hline
\end{tabular}
\caption{Examples of the hadron operators.
For baryon operators, the contraction with the color index is omitted.}
\label{tab:operator}
\end{table}

\paragraph{Baryon spectrum}

As listed in Table~\ref{tab:operator},
we use the standard baryon operators which have the same quantum
numbers as the corresponding baryons and survive in the nonrelativistic
limit.
At large $t$ (and large $N_t - t$), the baryon correlators are
represented as
\begin{eqnarray}
 G_B(t) \equiv  \sum_{\vec{x}} \langle
                      B(\vec{x},t) \bar{B}(\vec{x},0) \rangle
  &=&  \hspace{0.3cm}
    (1+\gamma_4) \left[ c_{B^+} \cdot e^{-tm_{B^+}}
           + b c_{B^-} \cdot e^{-(N_t - t) m_{B^-}} \right]
   \nonumber  \\
  & &
  + (1-\gamma_4) \left[ b c_{B^+} \cdot  e^{-(N_t - t)m_{B^+}}
             +  c_{B^-} \cdot e^{- t m_{B^-}} \right],
\end{eqnarray}
where $b=+1$ and $-1$ for the periodic and antiperiodic temporal
boundary conditions for the quark fields.
Thus, combining the parity-projected correlators under two
boundary conditions,
one can single out the positive and negative parity baryon
states with corresponding masses $m_{B^+}$ and $m_{B^-}$,
respectively.

\begin{figure}[tb]
\centerline{
\leavevmode\psfig{file=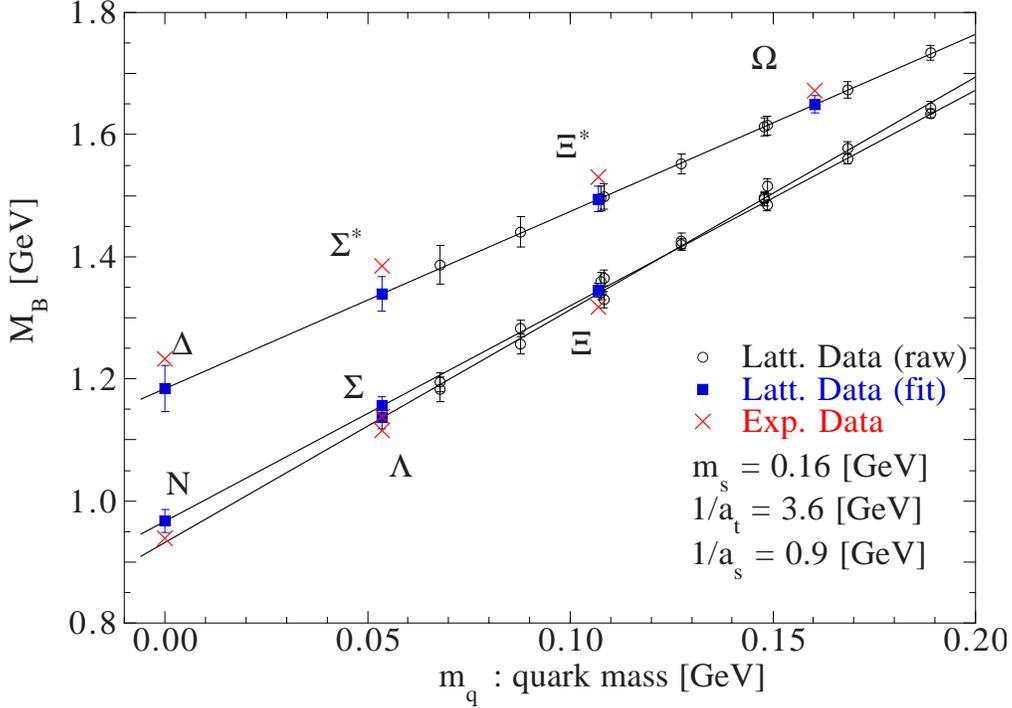,width=\figwidth}}
\caption{The spectra of octet and decuplet baryons with
positive parity. The horizontal axis denotes the averaged current quark mass.}
\label{fig:fig1}
\end{figure}

In our calculation, $u, d$ current quark masses are taken to be 
the same value as $m_u=m_d$ or $\kappa_u=\kappa_d$, and 
the strange current quark mass $m_s$ or $\kappa_s$ is 
taken to be an independent value. 
Then, the baryon masses are expressed as the function of $m_u$ and $m_s$ like 
$M_B(m_u,m_s)$, and therefore the baryon masses $M_B(m_u,m_s)$ are to be 
depicted on the $(m_u,m_s)$ plane. 
However, the lattice QCD result for the baryon masses $M_B$ seem to be well 
described as a function of the averaged current quark mass
over the three quarks, $m_q=(m_1+m_2+m_3)/3$, like $M_B(m_q)$
in each channel.
Therefore, we will use the simplified figures as the function of 
$m_q$, although the actual lattice QCD calculations are performed 
with different quark masses on $m_u (=m_d)$ and $m_s$. 

Let us start with the ground-state baryons shown in
Figure~\ref{fig:fig1}, where 
the horizontal axis denotes the averaged current quark mass over the three
quarks in the physical unit. 
We use the naive relation between the current quark mass and the
hopping parameter as 
\begin{equation}
 m_q = \frac{1}{3} (m_1 + m_2 + m_3 ), \hspace{1cm}
 m_i = \frac{1}{2}\left( \frac{1}{\kappa_i} - \frac{1}{\kappa_c}\right).
\end{equation}
The baryon masses are linearly extrapolated to the chiral limit, $m_q=0$.
In the analysis of the baryon spectrum, we use the lattice scale
determined by setting the averaged mass over the octet and the
decuplet baryons at the chiral limit to the averaged
mass of $N$ and $\Delta$.
It results in $a_{\sigma}^{-1}=0.9$ GeV.
The strange quark mass is determined so that the octet and the
decuplet baryons in Fig.~\ref{fig:fig1} globally reproduce
the experimentally measured masses.
Thus, for the strange quark, we adopt $\kappa_s = 0.1229(1)$, which roughly 
corresponds to $m_s \simeq 0.16$ GeV as the current quark mass. 
This value seems consistent with the standard strange current-quark mass. 
Hereafter, we will fix $\kappa_s = 0.1229(1)$ for the strange quark. 
Figure~\ref{fig:fig1} shows that the lattice result reproduces
that the measured baryon spectrum within 5 \% deviation.
  
\paragraph{Octet-Decuplet mass splitting}

Let us consider the difference of octet and decuplet baryon masses.
This mass splitting monotonously decreases 
as the current quark mass increases. 
This tendency is consistent with the one-gluon-exchange explanation
in the nonrelativistic quark model, assuming the constituent quark
mass $M_q$ increases with the current quark mass $m_q$, 
as $M_q \simeq M_0+m_q$.

\begin{figure}[tb]
\centerline{
\leavevmode\psfig{file=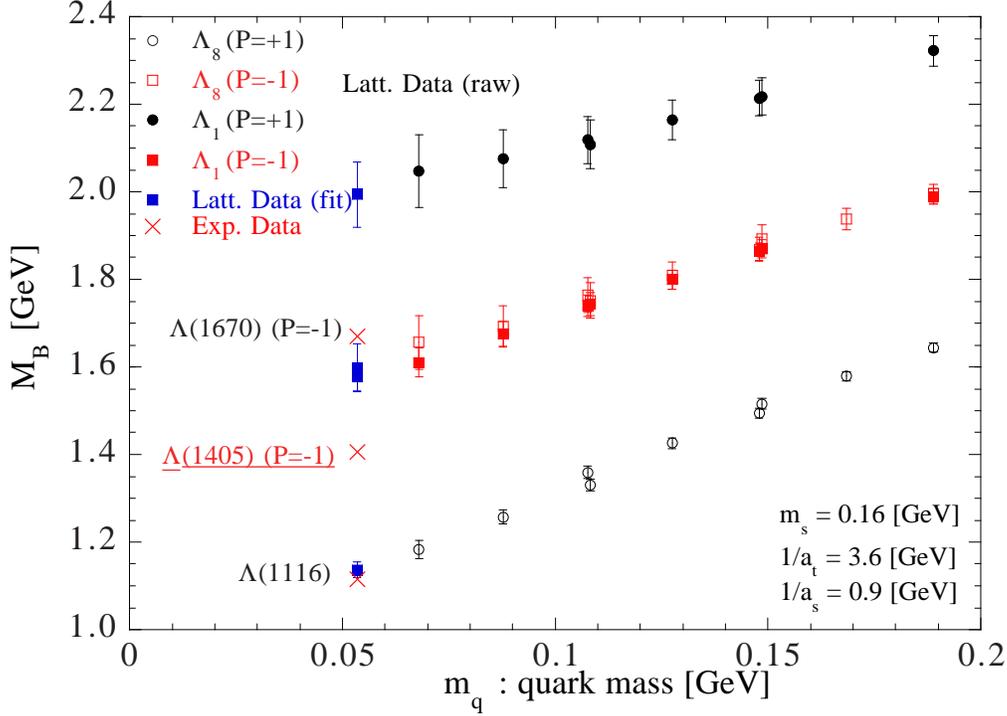,width=\figwidth} }
\caption{The positive and negative parity baryon masses
in the flavor octet and singlet channels.}
\label{fig:fig3}
\end{figure}

\paragraph{Negative parity baryons}

Figure~\ref{fig:fig3} shows the mass spectra of 
the flavor octet and singlet baryons
with positive and negative parities as the function of 
the averaged current quark mass $m_q$. 
In each channel, the baryon mass spectrum shows clear linear behavior 
on the averaged current quark mass $m_q$ in the whole measured mass region. 
Then, we extrapolate the masses to the physical situation 
$m_u=m_d \simeq 0, m_s=0.16 {\rm GeV}$, i.e., $m_q \simeq m_s/3$.
One striking feature is that the singlet and octet negative
parity baryons are almost degenerate.
On the other hand, the lowest mass of the flavor-singlet
positive-parity baryon is much larger than those in other channels. 
Extrapolating to the physical situation 
$m_u=m_d \simeq 0, m_s=0.16 {\rm GeV}$ ($m_q \simeq m_s/3$),  
we obtain 1.6 GeV for the lowest mass of the flavor-singlet 
negative-parity baryon, which is described as a three quark state 
in the quenched lattice.
Then, the experimentally observed negative-parity baryon 
$\Lambda(1405)$ is much lighter than the lowest mass of 
the flavor-singlet negative-parity baryon obtained in lattice QCD 
at the quenched level. 
Since the $q$-$\bar q$ pair creation is absent in the quenched QCD,
this result implies that the simple three valence quark picture
would not valid for $\Lambda(1405)$.
One possible explanation is, as frequently suggested, that
the $\Lambda(1405)$ is a  mixture of a three quark state and
an $N\bar{K}$ bound state.
To clarify this subject, we are going to measure the
$N\bar{K}$ state on the quenched lattice, where genuine
bound state properties is apparent owing to the absence of 
the $q$-$\bar q$ pair creation.

\section{Summary and outlook}
  \label{sec:Summary}

In this paper, we report present status of our investigation
of hadron properties using lattice QCD simulations.
At this stage, systematic investigation of baryon spectrum is 
in progress at the quenched level.
The result on the anisotropic lattice with $a_{\sigma}^{-1}\simeq 0.9$
GeV and anisotropy $\xi=4$ is as follows:
(a) The octet-decuplet baryon mass splitting decreases with increasing
current quark mass.
This tendency is consistent with the 
one-gluon-exchange explanation in the constituent quark model. 
(b) The negative-parity baryons in octet and singlet channels are
measured in good statistical precision. 
However, the experimentally observed negative-parity baryon $\Lambda(1405)$ 
seems much lighter than the lattice QCD result for the lowest mass of 
the SU(3) flavor-singlet negative-parity baryon. 
This may suggest the possibility of that $\Lambda(1405)$ 
is a mixture of the $N \bar K$ bound state and the three quark state.  
This would be clarified by successive lattice calculations 
for $qqq$-$q\bar q$ system in the quenched approximation.

Our present results have been carried out on rather course lattice. 
Hence, we need to perform the simulations on finer lattices
to remove lattice artifacts.
Then, the anisotropic lattice will be a powerful device to study
detailed properties of hadrons including excited states, exotics and
glueballs.
The simulation with dynamical quarks are also to be performed.

Another course of our program is a comparison with the potential
model analysis using the static quark potential extracted from
lattice QCD.
This will  give us important information on the quark wave
function and novel insight on the quark structure of hadrons.

We are grateful to Profs. Il-Tong Cheon and Su~Houng~Lee 
for their warm hospitality at Yonsei University. 
The lattice QCD simulations have been performed on NEC SX4 at Osaka
University, and Hitachi SR8000 at KEK.

\end{document}